\documentclass[sigconf,nonacm]{acmart}
\usepackage{listings}
\usepackage{tabularray}
\usepackage{stfloats}
\usepackage{enumerate}
\usepackage{enumitem}
\usepackage{xspace}
\usepackage{verbatim}
\usepackage{tablefootnote}
\usepackage{booktabs}
\usepackage{diagbox}
\usepackage{threeparttable}

\setlist[itemize]{leftmargin=3.2mm}

\newcommand{\toolname}[0]{QVoG\xspace}

\AtBeginDocument{%
  }

\begin{document}

\title{Scalable Defect Detection via Traversal on Code Graph}

\author{Zhengyao Liu}
\affiliation{
  \institution{Beihang University}
  \city{Beijing}
  \country{China}
}
\email{zhengyaoliu@buaa.edu.cn}

\author{Xitong Zhong}
\affiliation{
  \institution{Beihang University}
  \city{Beijing}
  \country{China}
}
\email{3012290812@qq.com}

\author{Xingjing Deng}
\affiliation{
  \institution{Fuzhou University}
  \city{Fuzhou}
  \country{China}
}
\email{micro6947@gmail.com}

\author{Shuo Hong}
\affiliation{
  \institution{Beihang University}
  \city{Beijing}
  \country{China}
}
\email{hongshuo@buaa.edu.cn}

\author{Xiang Gao}
\authornote{Corresponding author}
\affiliation{
  \institution{Beihang University}
  \city{Beijing}
  \country{China}
}
\email{xiang_gao@buaa.edu.cn}

\author{Hailong Sun}
\affiliation{
  \institution{Beihang University}
  \city{Beijing}
  \country{China}
}
\email{sunhl@buaa.edu.cn}

\begin{abstract}

Detecting defects and vulnerabilities in the early stage has long been a challenge in software engineering.
Static analysis, a technique that inspects code without execution, has emerged as a key strategy to address this challenge.
Among recent advancements, the use of graph-based representations, particularly Code Property Graph (CPG), has gained traction due to its comprehensive depiction of code structure and semantics.
Despite the progress, existing graph-based analysis tools still face performance and scalability issues.
The main bottleneck lies in the size and complexity of CPG, which makes analyzing large codebases inefficient and memory-consuming.
Also, query rules used by the current tools can be over-specific.
Hence, we introduce \toolname, a graph-based static analysis platform for detecting defects and vulnerabilities.
It employs a compressed CPG representation to maintain a reasonable graph size, thereby enhancing the overall query efficiency.
Based on the CPG, it also offers a declarative query language to simplify the queries.
Furthermore, it takes a step forward to integrate machine learning to enhance the generality of vulnerability detection.
For projects consisting of 1,000,000+ lines of code, \toolname can complete analysis in approximately 15 minutes, as opposed to 19 minutes with CodeQL.

\end{abstract}





\maketitle

\section{Introduction}

In today's software engineering, effectively detecting defects or vulnerabilities in the early stages of development remains a challenge.
The later they are discovered, the higher the cost of repair \cite{Zheng2006}. 
To address this issue, static analysis has long been a useful technique \cite{Emanuelsson2008}.
It requires no dynamic information but can provide satisfying results for possible vulnerabilities in the code.

In recent years, there has been a growing interest in vulnerability detection using graph queries.
The main concept involves using the Code Property Graph (CPG) to comprehensively represent the source code \cite{Yamaguchi2014}.
Since vulnerable code often follows specific patterns, the detection process can be translated into graph queries.
Static analysis tools that rely on graph queries typically operate in two stages: extracting the CPG from the source code and then executing queries on the graph.
Joern \footnote{\url{https://joern.io/}} \cite{Yamaguchi2014} and cpg \footnote{\url{https://github.com/Fraunhofer-AISEC/cpg}} \cite{Weiss2022} are two examples that make use of CPG for vulnerability detection.

\subsection{Challenges}

Although the current detection tools and methods based on graph query analysis have yielded good results, there are still challenges that need to be addressed.

\begin{itemize}
    \item Most of the existing code analysis tools construct the CPG at the AST level, which results in an extremely huge graph as the codebase grows.
    Complex CPG may impact the performance of the analysis.
    \item Existing query language is complex and difficult to get started.
    For example, Joern uses a Scala-based domain-specific query language\footnote{\url{https://docs.joern.io/}},  which may increase the complexity of writing custom queries for users.
    \item Analysis for large projects requires more resources and time.
    Memory usage is the main resource concern, and machines with low RAM may be unable to perform large-scale analysis.
    So, more attention is needed to the scalability of analysis on large projects.
    \item Detecting vulnerability via query rules can be over-specific. Such queries may be accurate but lose generality for similar problems, hence resulting in false positives or negatives.
\end{itemize}

\subsection{Overall Methodology}

To tackle these challenges, we propose a general approach, \toolname, for detecting defects and vulnerabilities in large software systems based on graph query analysis.
\toolname aims to provide a complete analysis process from the CPG extraction to the graph query.
The main components consist of CPG extraction and a query engine that executes vulnerability queries.

More specifically, to reduce the graph complexity in traditional CPG, we propose a novel structure of CPG with necessary information only and much smaller.
It uses a single statement node to replace all its Abstract Syntax Tree (AST) nodes, making the latter as attributes.
This way can reduce the traverse steps for graph query thus improving the overall performance.
With ease of use in mind, we designed a declarative domain-specific language for graph queries to make it simpler to write specific query rules.
We have also taken into account language differences when developing the query engine to ensure seamless support for new programming languages with minimal effort.
For scalability, we incorporate multiple optimizations throughout the workflow to improve efficiency on large-scale projects.

To alleviate the over-specific problem, we utilize machine learning to achieve a more general detection of vulnerabilities.
By training our models with existing datasets and integrating them with the query engine, we enhance the generalization ability of vulnerability queries.

\subsection{Evaluation}

To measure the efficiency of \toolname, we compare with graph-based Joern, and non-graph-based CodeQL \footnote{\url{https://codeql.github.com/}}, and run all three tools on the same datasets with semantically equivalent queries.
Based on this, we make evaluations mainly on the following metrics.

\begin{itemize}
    \item \textbf{Query Accuracy} --- Precision and recall of defect detection on common CWE vulnerabilities.
    \item \textbf{Performance and efficiency} --- Time and memory cost for CPG or database (for CodeQL) extraction and query execution.
\end{itemize}

The results show that \toolname has a reasonable time and memory cost for analysis on both small and large projects.
For a project with more than 1,500,000 lines of code, \toolname can complete CPG extraction in approximately 15 minutes compared with 19 minutes of CodeQL, with memory cost much lower than Joern.
As for accuracy, we evaluate \toolname on Juliet test suites and it demonstrates a 90\% precision and 95\% recall rate on average, surpassing that of both Joern and CodeQL.
Moreover, compared to Joern and CodeQL, \toolname will be fully open source.

\subsection{Contributions}

The contributions of this article can be summarized as follows:

\begin{itemize}
    \item \textbf{Compressed Code Property Graph} Conventional CPG includes many redundant graph nodes and edges that impact the performance of the analysis. 
    Hence, we propose a novel structure of CPG containing all necessary information yet smaller.
    \item \textbf{Dedicated Domain Specific Language} We design a DSL suitable for graph queries with a syntax similar to SQL.
    It allows user to write their queries simpler and much easier.
    \item \textbf{Language-independent query interface} A consistent query interface could help scale the capability for more programming languages.
    Therefore, we introduce a language-independent code representation with a set of fluent APIs as the foundation of vulnerability detection.
    \item \textbf{Combination of graph query and deep learning} We integrate the logical reasoning of graph query and the learning ability of deep learning, the ultimate goal is to enhance the generalization of the pre-defined queries and increase detection accuracy.
    \item \textbf{Open-source Tool} Different from CodeQL, whose query engine is close source, and Joern, whose inter-procedure analysis is close source, \toolname will be made fully open-source.
\end{itemize}

\section{Methodology}

We developed a graph-based static analyzer called \toolname, a general platform for detecting defects and vulnerabilities.
We aim to improve the efficiency of CPG extraction and code query and achieve an efficient, accurate, and extensible CPG framework.

\subsection{Key Concepts}

To fully support the complete analysis workflow, \toolname encompasses functions from CPG extraction to query execution.
Before delving into the details of each part, it's important to introduce several key concepts.

\paragraph{Code Property Graph}

Classic static analysis involves AST, Control Flow Graph (CFG), Program Dependency Graph (PDG), and other flow information of the source code.
To combine these pieces of information, CPG was introduced as a novel code representation \cite{Yamaguchi2014}.
Based on CPG, common types of vulnerabilities can be detected by performing certain traversals on the graph.
To avoid the redundancy of AST nodes, we compress the structure of CPG to the statement level with AST information as attributes.
This helps to keep the entire graph at an acceptable size as the project grows.
Besides, we use Data Flow Graph (DFG) instead of PDG to depict more detailed data dependency.

\paragraph{Domain-Specific Language Design}
\label{section:dsl}

For tasks in a specific area, Domain-Specific Language (DSL) can provide users with a simpler and easier way to interact with the underlying system. \cite{Karsai2014}
Declarative DSL like SQL also hides the concrete implementation from users, so that they can focus on the query.
To achieve a more user-friendly query interface, we designed a declarative DSL dedicated to graph queries.
Behind the DSL, we use a set of fluent APIs to support a consistent interface and calling convention. \cite{Roth2023}

\paragraph{Query Execution And Optimization}

The query will first be written in DSL, then a translator will first convert it into our query API, and eventually into Gremlin Query Language to query data from the graph database.
Database access is expensive, especially as the graph size increases.
Therefore, we introduce parallel computing and caching during graph query to improve \toolname's efficiency.

\paragraph{Combination of Query And Machine Learning}

The rule-based query is efficient at detecting known defects, but it suffers from over-specific issues.
In this case, we propose a combination of query and machine learning to detect more variants of defect patterns.

\subsection{Architecture}

\begin{figure}[t]
  \centering
  \includegraphics[width=\linewidth]{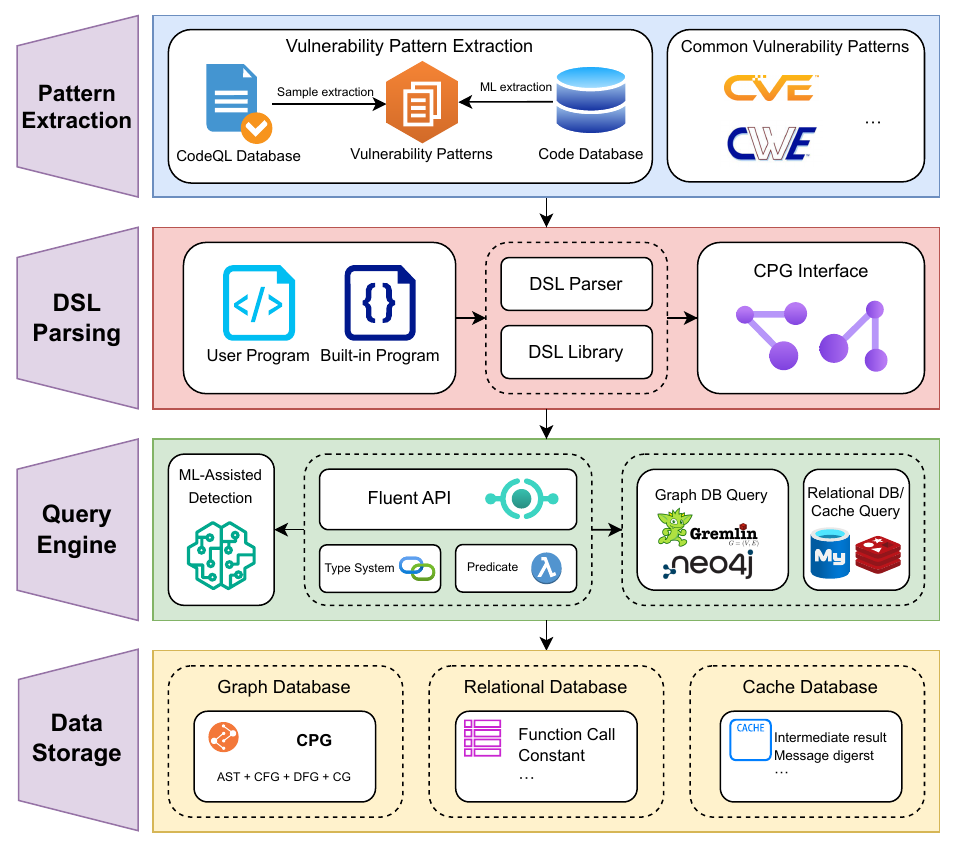}
  \caption{Architecture of \toolname}
  \Description{The overall architecture of \toolname with four modules.}
  \label{fig:Architecture}
\end{figure}

In this section, we will present the overall architecture of \toolname.
As shown in Figure \ref{fig:Architecture}, it involves four modules, each providing a layer of support for vulnerability detection.
More details for each module will be discussed in the following sections.

\toolname builds CPG for the source code using static analysis tools for different languages, including Java, Python, C, etc.
On top of the CPG, we designed a DSL for users to write queries for vulnerability detection.
The design considered the defects and vulnerability patterns from public datasets like CWE \footnote{\url{https://cwe.mitre.org/}} and CVE \footnote{\url{https://www.cve.org/}}.
During the execution of queries, we combine the formal constraints of DSL and probabilistic machine learning prediction to improve the generality of extracted patterns.
Finally, different types of code information are stored in suitable databases to improve data access efficiency.
Specifically, CPG is stored in a graph database, while function information and other summarized characteristics are in a relational database.
Moreover, caching is used for temporary results to speed up the query further.

\begin{figure}[!th]
  \centering
  \includegraphics[width=\linewidth]{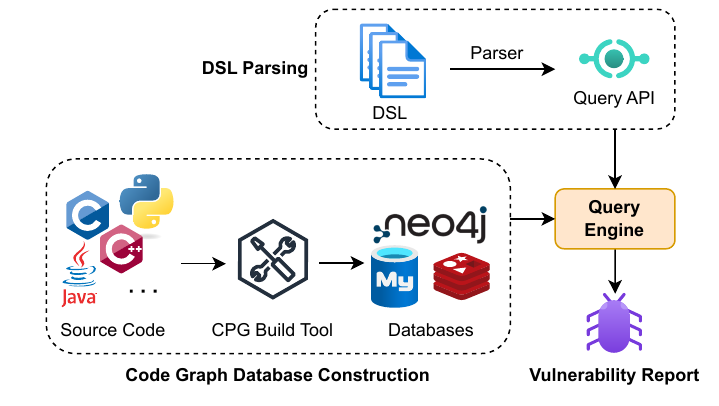}
  \vspace{-6pt}
  \caption{Workflow of \toolname}
  \Description{The two-stage workflow of \toolname.}
  \label{fig:Workflow}
  \vspace{-6pt}
\end{figure}

We divide the analysis process into two decoupled stages --- front-end CPG extraction and back-end query execution.
As is shown in Figure \ref{fig:Workflow}, \toolname first extracts CPG and other information from the source code and stores them in corresponding databases.
Then, the query engine executes the user's queries on the databases and finally outputs a vulnerability report.
Since the two stages are decoupled, each stage can be more focused and flexible.

\subsection{Compressed Code Property Graph}

As proposed by Yamaguchi et al.~\cite{Yamaguchi2014}, the CPG is a graph-based representation of source code that captures the structure and semantics of the code.
Specifically, CPG is a directed graph where nodes represent code entities and edges represent the relationships between them.
It combines AST, CFG, PDG, and Call Graph (CG).
AST captures the syntactic structure of the code, while the CFG and CG capture the control flow structure.
PDG captures the control and data dependency.
By combining these four graphs, we can capture the structure and semantics of the code.
CPG can be used for various static analysis tasks, including vulnerability detection, code comprehension, and program transformation.

The conventional CPG(Joern) is stored at the level of AST nodes, which may lead to too many nodes and edges as the code base grows.
Given the following piece of code in Listing \ref{lst:CPG-Example} as an example.
The graph includes about 100 nodes and 460 edges when using Joern to extract the complete CPG.
Such a high number of nodes and edges could impact the performance of graph analysis as traversal overhead increases.

\begin{lstlisting}[language=c, caption=Example C program to extract CPG, label=lst:CPG-Example]
#include <stdio.h>
int main(void) {
    int a = 2;
    int b = a * a;
    if (b > a) {
        b = b - a;
    }
    printf("a + b = %d\n", a + b);
    return 0;
}
\end{lstlisting}

To alleviate this issue, we propose a compressed CPG that stores the code at the level of statements.
Instead of storing AST nodes directly in the graph database, we make them attributes of the statement they belong to.
Our CPG can be defined as $G = (V, E)$, where $V$ is the nodes and $E$ is the edges.
$V$ represents a statement in the source code, its attributes are shown in the following list.
It mainly contains the location and AST structure of the statement.
Other properties such as function call information will also be included as summary information.

\begin{itemize}
    \item $file$: The full path of the file this statement belongs to.
    \item $lineno$: The line number of this statement in the file.
    \item $code$: The original code of the statement.
    \item $ast$: The minimum AST in JSON format.
    \item $call$: (optional) Function name if the statement is a function call.
\end{itemize}

For $E$, it represents the flow between statements.
In our implementation, we mainly combine CFG, DFG, and CG to build the CPG.
So edges are labeled with the corresponding flow types such as $CFG$, $DFG$, and $CG$.
With statement-level CPG, the code in Listing \ref{lst:CPG-Example} only yields about 10 nodes and 15 edges.
By simplifying the structure of CPG, we can improves the overall efficiency of queries.

While statement-level CPG eliminates most redundancy, it also loses some detailed information, especially in DFG.
DFG can start and end from exact variables in AST in a complete CPG representation.
However, in our compressed CPG, DFG can only reach statements and therefore cannot accurately determine the actual data flow when multiple data dependencies are involved.
To address this issue, we annotate the DFG edge with the corresponding data dependency.
Such improvement can help to recognize complex data dependency and make defect detection more precise.

Moreover, it's common to have variable aliasing, where multiple variables point to the same memory location.
This adds complexity to data flow analysis because changes to one variable affect all variables that point to this location.
\toolname uses alias analysis to record variables created through assignments as aliases and embed the results in DFG.
This improves the precision of data flow analysis and helps us better understand complex program structures.

\subsection{Declarative DSL for Graph Query}

Declarative DSL focuses on what needs to be done, not how to do it.
Therefore, we design a declarative DSL for graph query that is specialized for querying the compressed CPG.
The query operators are designed to be simple and intuitive so that users can easily write queries.
The extended Backus–Naur form grammar of the DSL is shown in Listing \ref{lst:DSL}.

\begin{figure*}
\begin{lstlisting}[language=, caption=BNF grammar of the DSL, label=lst:DSL]
Q    := from <decl> {, <decl>}
        [where [not] <predicate> {and|or [not] <predicate>}]
        select <expr> {, <expr>}
<decl> := <type> name | <predicate> name
<expr> := <name> | <string>
\end{lstlisting}
\end{figure*}

The operators are similar to that of SQL, but the underlying working mechanism is different.
Specifically, the \lstinline{from} clause is used to specify one or more sets of nodes in the graph as query context.
Using \lstinline{decl}, users can specify the type of nodes to query.
A predicate can also be used for a more precise context.
The \lstinline{where} clause is used to specify one or more predicates to filter the context.
It mainly uses the flow information to check whether a path exists between the given nodes.
It can also filter the nodes with more predicates.
Finally, the \lstinline{select} clause is used to specify the nodes to return in the query result.
An example of a DSL program is shown in Listing \ref{lst:DSLExample}, which is used to detect simple code injection vulnerability.
It selects the call nodes that call the \lstinline{input} function as source and the \lstinline{exec} function and sink.
Then, it finds if there is a taint flow from the source to the sink.

\begin{lstlisting}[language=, caption=DSL for code injection, label=lst:DSLExample]
from Call a, Call b, TaintFlow flow
where
  a.getFunction().equals("input") and
  b.getFunction().equals("exec") and
  flow.source(a).sink(b).exists()
select a, b, flow
\end{lstlisting}

\subsection{Query Engine}

\begin{figure}[!th]
  \centering
  \includegraphics[width=\linewidth]{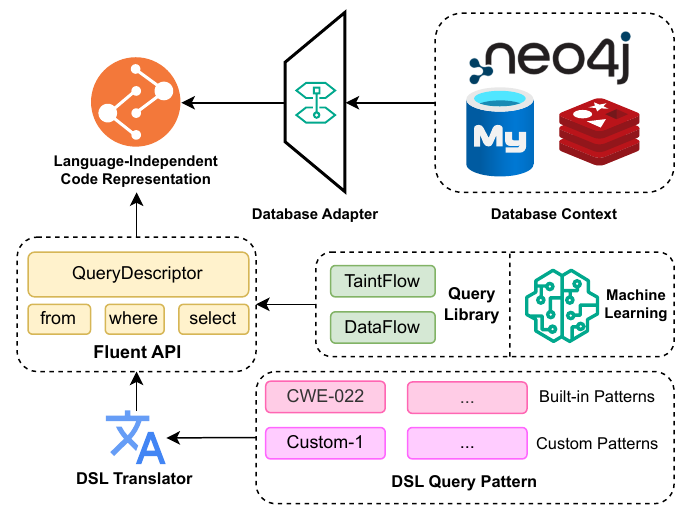}
  \vspace{-6pt}
  \caption{Query engine architecture}
  \Description{The overall architecture of the query engine.}
  \label{fig:QueryEngine}
  \vspace{-6pt}
\end{figure}

The overall architecture of the query engine is shown in Figure~\ref{fig:QueryEngine}.
The query engine is the backend to execute the DSL on the CPG.
It consists of four parts --- language-independent code representation, database adapter, query API, and DSL translator.
The query engine first fetches the CPG information from the database.
With the help of the database adapter, the CPG is converted into a language-independent representation.
It shields the differences between programming languages to have a consistent query interface.
Based on this, a set of Fluent API is provided to support query operations on the CPG.
Patterns written in DSL are then translated into query APIs and then executed to query database.

\begin{figure*}[!th]
\begin{lstlisting}[language=java, caption=Query API example for code injection, label=lst:FluentAPIQuery]
QueryDescriptor.open()
    .from("a", CallExpression.class)
    .from("b", CallExpression.class)
    .where(q -> q.onTable("a").where(new ContainsFunctionCall("input")))
    .where(q -> q.onTable("b").where(new ContainsFunctionCall("exec")))
    .where(TaintFlowPredicate.with().source("a").sink("b").as("flow").exists())
    .select("a", "b", "flow");
\end{lstlisting}
\end{figure*}


\subsubsection{Language-Independent Code Representation}

To mask differences between languages, an intermediate code representation is often used.
It makes the analysis tool more flexible, and scalable for new programming languages. \cite{Weiss2022}
The language-independent code representation is designed based on our CPG. 
The ideas are inspired by LLVM IR \footnote{\url{https://llvm.org/}} and cpg \footnote{\url{https://github.com/Fraunhofer-AISEC/cpg}}.
This representation is used for a unified AST structure to represent detailed information of statements.
A strong type system is used to ensure well-formatted code structure and provide interface with combined language features.

\subsubsection{Database Adapter}

\begin{figure}[h]
  \centering
  \includegraphics[width=\linewidth]{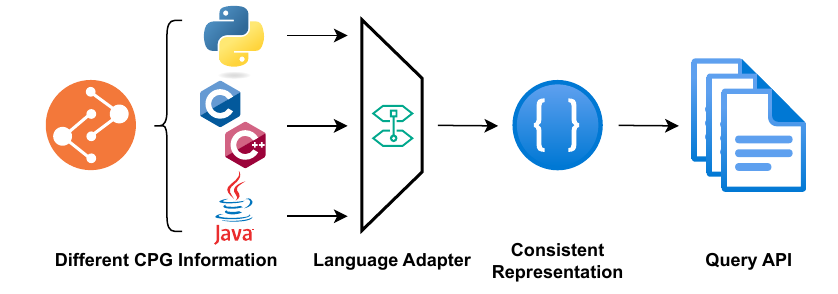}
  \vspace{-6pt}
  \caption{Database adapter workflow}
  \label{fig:Adapter}
  \vspace{-6pt}
\end{figure}

Nevertheless, not all differences between programming languages, especially AST structures, can be avoided.
For example, the syntax of function calls in C is different from that in Python, which results in different AST node types and structures.
To address this issue, an adapter is implemented to convert the AST of different programming languages into the language-independent code representation defined above.
The adapter is designed to be extensible so that users can easily add support for new programming languages.
Its workflow is shown in Figure \ref{fig:Adapter}.

Using the $ast$ property stored in CPG, we can re-construct the AST and convert it into our language-independent representation.
This conversion happens in the adapter, which is transparent to the query engine.
So that the query engine can perform the execution regardless of the underlying programming language.

\subsubsection{Query API And DSL Translator}

The user-written DSL will eventually be translated into Gremlin Query Language (Gremlin) to traverse the graph database.
However, the steps provided by Gremlin are for general purposes only and thus not suitable for querying code properties.
Therefore, we add an extra layer between our DSL and Gremlin to enable complex queries on the code graph.
Based on our CPG structure, we adopt the idea of fluent API to provide a consistent query interface \cite{Roth2023}.
It provides a straightforward way to translate DSL into query API and perform type checking and error handling in compile-time.
Listing \ref{lst:FluentAPIQuery} is an example of the query API, translated from the DSL given in Listing \ref{lst:DSLExample}.
Benefiting from the syntax of fluent API, we can translate the DSL quite straightforwardly.

\subsection{Machine Learning}
To solve the over-specific issues related to DSL-based rules, a machine learning component is introduced to define more general query patterns.
As the first step, we employ machine learning for taint analysis, which is the most widely used pattern in defect detection.
The fundamental concept of taint analysis involves designating certain data as tainted Sources, followed by tracking the propagation paths of tainted data throughout the program until they are utilized in sensitive operations, i.e., Sinks.

\begin{figure}[!th]
  \centering
  \includegraphics[width=\linewidth]{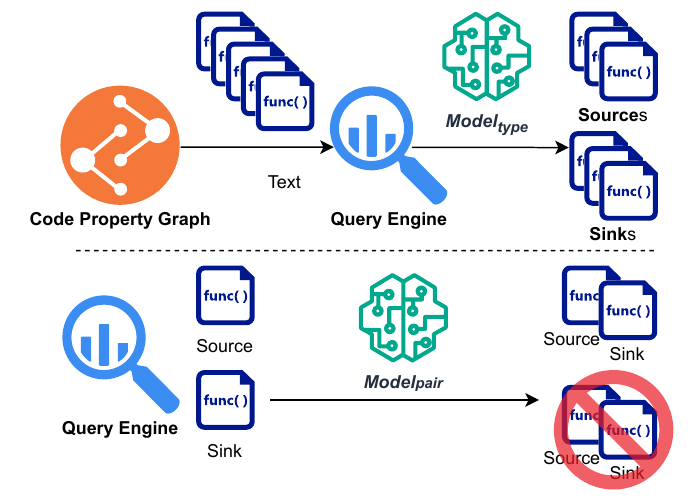}
  \vspace{-6pt}
  \caption{Workflow of Query Engine Combining Models}
  \label{fig:MLArchitecture}
  \vspace{-6pt}
\end{figure}

Specifically, the workflow is depicted in the figure \ref{fig:MLArchitecture}.
To identify tainted Sources and vulnerability-prone Sinks, a model ($Model_{type}$) is trained to classify each line of code.
However, straightforward classification will lead to false positives, as identified sources and sinks on a control flow may not correspond to the same vulnerability type. 
Therefore, a second model ($Model_{pair}$) is trained to verify whether a given Source-Sink pair matches correctly, refining the analysis and reducing false alarms.
When the query engine operates, it initially submits the nodes of the CPG to $Model_{type}$ for analysis, which determines the classification of each node. 
The engine then starts from the Source nodes and traverses downstream along the control and data flows.
When encountering a node classified as a Sink, $Model_{pair}$ is used to check if the Source and Sink match. 
If they match, it indicates a potential vulnerability.

We train our models by fine-tuning CodeBERT \cite{Feng2020}. 
The reason for adopting CodeBERT lies in the distinctive syntax and semantic structure found in code, which is quite different from natural language and unsuitable for applying traditional natural language processing models.
As CodeBERT is pre-trained on extensive code datasets, it grasps the specific syntax, structure, and constraints inherent in code.

\begin{figure}[h]
  \centering
  \includegraphics[width=0.8\linewidth]{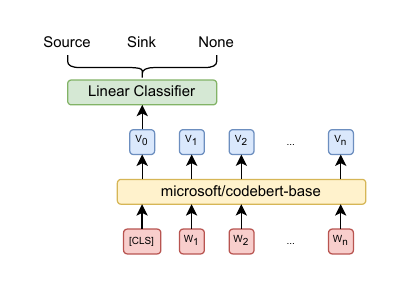}
  \vspace{-6pt}
  \caption{$Model_{type}$ architecture}
  \vspace{-6pt}
  \label{fig:Type_model}
\end{figure}

\paragraph{$Model_{type}$ Architecture}

The architecture of $Model_{type}$ is shown in Figure \ref{fig:Type_model}. 
In this diagram, ``microsoft/codebert-base'' indicates the pre-trained model, with the [CLS] token summarizing the overall semantics of the code snippet. 
$W_1$ to $W_n$ represent the tokens obtained after tokenizing a line of code.
The CodeBERT model processes the [CLS] token and the token list, producing an output vector list, $V_0$ to $V_n$. 
Here, $V_0$ captures the comprehensive meaning of the statement. 
Passing this vector through a linear layer and taking the maximum value allows the model to perform trichotomous classification.


\begin{figure}[!th]
  \centering
  \includegraphics[width=\linewidth]{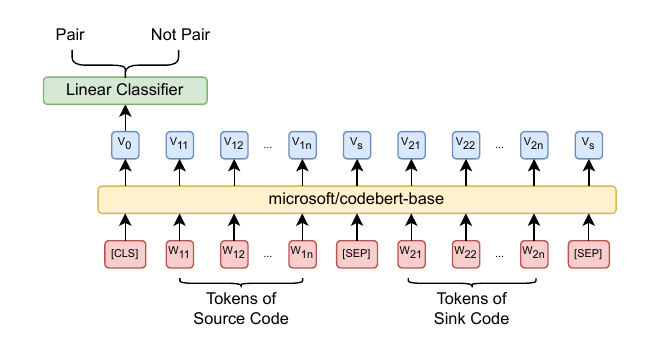}
  \caption{$Model_{pair}$ architecture}
  \label{fig:Pair_model}
\end{figure}

\paragraph{$Model_{pair}$ Architecture}


Identifying Sources and Sinks is insufficient without determining if they form valid pairs. Thus, we train $Model_{pair}$ for this purpose.
The architecture of $Model_{pair}$ is shown in Figure \ref{fig:Pair_model}. 
$Model_{pair}$ is similar to $Model_{type}$ but differs in input format and classification type. 
In specific, $Model_{pair}$ follows BERT's input format, taking token lists representing two lines of code, separated by [SEP] tokens at the start and end. 
For classification, $Model_{pair}$ handles a binary task to determine if a Source and Sink form a pair, outputting 0 for non-pairing and 1 for pairing.


\section{Implementation}

In this section, we provide a general overview of how our approach is implemented.
The implementation mainly involves CPG extraction, a query engine, and flow analysis.
Additionally, we have applied multiple optimization strategies to further enhance the query performance.

\subsection{CPG Extraction}

The primary focus of our approach is on extracting CPG information.
While there are several analysis tools available for different programming languages, most of these tools only parse the source code and produce AST.
Therefore, we have extended these tools to extract CPG information.
Currently, \toolname supports C, Python, and Java languages.
The analysis tools used for each language are listed in Table \ref{tab:AnalysisTool}.

\begin{table}
    \centering
    \caption{Analysis tool used for CPG extraction}
    \label{tab:AnalysisTool}
    \begin{tabular}{cc}\toprule
        \textbf{Programming Language} & \textbf{Analysis Tool Used} \\
        \hline
        C & Eclipse CDT \tablefootnote{\url{https://projects.eclipse.org/projects/tools.cdt}} \\
        Python & Scalpel \cite{Li2022} \\
        Java & Eclipse JDT \tablefootnote{\url{https://projects.eclipse.org/projects/eclipse.jdt}} \\
        \bottomrule
    \end{tabular}
\end{table}

As the CPG extraction is decoupled, this process can be more flexible, utilizing various techniques as long as the output CPG format meets our definition.
This flexibility enhances the analysis and opens the door for potential technological upgrades and functional extensions in the future.
The extracted CPG will then be stored in a graph database.
To ensure maximum portability, we use the standard interface provided by Apache TinkerPop\texttrademark\space \footnote{\url{https://tinkerpop.apache.org/index.html}} framework.
Based on this, we choose Neo4j \footnote{\url{https://neo4j.com/}} as our graph database for its performance and reliability. \cite{Sholichah2020}
Finally, we embed Gremlin as the graph traversal language to store or retrieve graph information.

\subsection{Query Library}
\label{querylib}

One of the advantages of graph-based analysis is that it depicts structured flow information in code, and we integrate this analysis into a query library.
In this section, we discuss how to utilize the flow information in CPG to better detect defects and vulnerabilities.

\subsubsection{Taxonomy of Defect and Vulnerability}

Our in-depth study of various vulnerabilities has revealed that these issues can be categorized into two main types.
The first type consists of problems that stem directly from the code itself, often due to incorrect API usage.
This typically requires utilizing information from AST to directly identify issues through a close examination of the syntax or API usage.
The second type of vulnerability requires identification through a process known as propagation analysis, which involves examining how different sections of the code are linked and interact with each other.
When dealing with this type, it's essential to utilize the flow information between different statements in the code.
We can achieve this by thoroughly traversing the graph structure and matching specific patterns to identify potential defects, which is called flow analysis.

\subsubsection{Flow Analysis}

The integrated approach using CFG, DFG, and CG offers a powerful tool for complex program analysis, particularly in identifying and resolving complex vulnerabilities involving multiple functions and modules.

Using the common example of tainted data propagation, this issue often occurs when a program directly executes logic based on untrusted user input.
It involves three key concepts, Source, Sink, and Barrier (or Sanitizer).
Source is the unsafe input, Sink is the use of it, and Barrier is the validation.
If Source can reach Sink through a path without Barrier, then it indicates that there is an unsafe use of data.
In this case, we first identify possible Sources, which are typically user inputs such as forms, API requests, client parameters, etc.
Then, we track the data flow of the Sources to find possible unsafe use of them.
After this, we track the control flow from Source to Sink to see if there exists any Barrier that validates or closes the unsafe data.
This way, we can find potential defects in the code of this pattern.

\subsubsection{Accuracy Improvement}

While DFG is useful for tracking data flows within a program, it may not fully capture the relationships between global variables when used to pass data across functions. 
To overcome the limitations of traditional DFG, we have made some adjustments to its edges.
For each variable, an edge in the DFG is created from its declaration to its specific location of use.
This design allows us to trace back every variable used in a statement to its declaration.
To detect DFG relationships, we use a backward propagation method.
Initially, we identify the declaration locations of each involved variable through backward propagation.
Then, by evaluating declaration locations, we determine whether the operations derive from the same variable.
This approach improves the precision when analyzing variable usages and enhances the \toolname's ability to handle complex data relationships.

\subsection{Query Optimization}

When working on large projects with million lines of code, analyzing performance can be quite challenging.
Therefore, optimization is crucial to ensure reasonable time costs.
We primarily utilize parallel computing and caching to enhance overall performance. 
Additionally, we employ bulk operations when accessing databases to minimize data transfer overhead.

\subsubsection{Parallel Optimization}

When extracting CPG, analysis for each source file can be parallelized.
This allows for much-improved performance.
We first extract the CFG and DFG from each file in parallel, and then we extract the cross-file CG.
The upsertion to the database is also multi-threaded and paralleled with CPG computing to improve I/O throughput.

\subsubsection{Bulk Operations}

Graph databases, e.g., Neo4j, provide strong support for transactions and bulk operations.
This ensures that concurrent operations do not compromise data consistency and can enhance overall throughput.
Bulk operations allow for combining the upserting of individual nodes or edges to reduce network overhead, leading to substantial improvements in performance.

\subsubsection{Query Cache}

Many defect patterns produce the same intermediate results.
For example, a function call is very likely to be the source of a pattern.
Therefore, caching these results can speed up the following queries for the graph database.
When performing a query, its intermediate results will be stored automatically and shared with other queries on the same codebase.

\subsection{Model Training}

By collecting various datasets containing examples of Source and Sink, we can train the $Model_{type}$ and $Model_{pair}$ effectively.

\subsubsection{Dataset Generation}
To build the dataset required for training our models, we extensively surveyed various databases, ultimately focusing on the GitHub Advisory Database \footnote{\url{https://github.com/advisories}} and the sample queries within the QL library of CodeQL \footnote{\url{https://github.com/github/codeql}}.
Combining data from these sources, we created a labeled dataset.
We remove indentations to prevent classification errors based on formatting.

Finally, we have gathered more than 900 data entries (Named Basic Dataset) using a combination of scripting and manual processing.
Each entry contains a CWE identifier, the programming language utilized, and code snippets representing Source and Sink nodes.
The code snippets contain the code segments between Source and Sink nodes to provide contextual extracts.

For the $Model_{type}$ dataset, each data entry consists of a line of code accompanied by a label (i.e., Source, Sink, or None). 
We maintain a data ratio approximately at 1:1:5 (Source to Sink to None) to ensure balance and enhance model accuracy.
Regarding the $Model_{pair}$ dataset, each data entry includes a line of Source, a line of Sink, and a corresponding label (i.e., true or false). 
The data ratio is approximately 1:4 (true to false) to balance the dataset and improve the model's effectiveness in identifying correctly paired Source-Sink relationships.

\subsubsection{Training Settings}

In the selection of the model optimizer, we employed the AdamW optimizer, setting the learning rate to a commonly used value of $10^{-5}$ and epsilon to $10^{-8}$. 
Such configuration facilitated faster model convergence and yielded favorable training outcomes.

Initially, we attempted to employ the CodeBERT model directly without fine-tuning for classification (i.e., pre-training).
However, the results were suboptimal, so we proceeded to fine-tune the CodeBERT model using our datasets (i.e., post-training). 
During post-training, we split the dataset into an 80\% training portion and a 20\% test portion. 
For the three-class classification task of $Model_{type}$, we assess the performance using accuracy and the Kappa score, which are indicative of model reliability and agreement beyond chance. 
On the other hand, for the binary classification task in $Model_{pair}$, we rely on the accuracy and the F1 score, with the latter being particularly sensitive to the balance of true positive and true negative predictions. 


\subsubsection{Model Server Interface}
To enable the query engine to leverage the trained models while maintaining low coupling, we deploy the models on a server, where we create a Web API using the Flask framework \footnote{\url{https://flask.palletsprojects.com/}}. 
This design allows the query engine to interact with the models simply by invoking API calls.


\section{Evaluation}

To demonstrate the effectiveness of our approach, we conducted evaluations in comparison with Joern and CodeQL to answer the following research questions.

\begin{itemize}
    \item \textbf{RQ1}: What is the effectiveness of \toolname comparing to existing tools?
    \item \textbf{RQ2}: Is the declarative DSL easy to use?
    \item \textbf{RQ3}: What is the efficiency of CPG extraction?
    \item \textbf{RQ4}: Can machine learning alleviate the issues caused by over-specific rules?
\end{itemize}

All experiments were conducted on a Windows PC with a 2.6GHz Intel Core i7-9570H CPU and 16 GB of memory.
Due to the environment incompatibility, we use the host Windows machine to run \toolname, a VMware virtual machine (Ubuntu 20.04 with 4 CPUs and 10GB of memory) to run Joern (2.0.161). For CodeQL (2.17.2), we run it on both the host machine and WSL (Ubuntu 18.04).


\subsection{RQ1: What is the effectiveness of \toolname comparing to existing tools?}
\label{rq1}

To evaluate the results of the query library mentioned in section \ref{querylib}, we used the widely recognized Juliet test suites \footnote{\url{https://samate.nist.gov/SARD/test-suites/112}} to thoroughly assess \toolname.
This test suite contains test cases in various programming languages and is designed to evaluate the effectiveness of static analysis tools in identifying different security vulnerabilities.
Focusing on C language, we conducted a detailed analysis of three common types of defects as listed below.

\begin{itemize}
    \item \textbf{CWE-401 Memory Leak} The program fails to properly release allocated memory, leading to continuous memory consumption during its execution.
    \item \textbf{CWE-415 Double Free} The program attempts to free the same memory block twice, which can lead to program crashes or security vulnerabilities.
    \item \textbf{CWE-416 Use After Free} The program continues to use a memory block after it has been freed, which can result in unpredictable behavior or program crashes.
\end{itemize}

We compared \toolname with Joern and CodeQL, and the results are shown in Table \ref{tab:juliet}.
The data in the table is in the form of Precision/Recall Rate, which can be defined as follows.

\vspace{-3pt}
\[
\textbf{Precision} = \frac{\text{Positives}}{\text{Positives} + \text{False Positives}} \times 100\%
\]
\[
\textbf{Recall Rate} = \frac{\text{Positives}}{\text{Total Vulnerabilities}} \times 100\%
\]

Note that, for \toolname and CodeQL, we report two sets of results, representing ``must'' and ``maybe'' results respectively.
The ``must'' only reports detected bugs with high confidence, while the ``maybe'' reports all the detected results.
Since Joern's query library currently does not include queries for CWE-401 and CWE-415.
Therefore, we write these queries by ourselves.


\begin{table*}
    \caption{Verification on Juliet test suites}
    \vspace{-6pt}
    \label{tab:juliet}
    \begin{tabular}{c|ccc}
        \toprule
        \diagbox[width=10em]{CWE Type}{Tool Name} & \toolname & Joern & CodeQL \\
        \hline
        CWE-401 &  92.50\%/66.07\% (61.53\%/\textbf{100\%})  &  77.08\%/66.07\%  &   \textbf{96.96}\%/57.14\% (92\%/41.07\%)  \\
        CWE-415 &  85.36\%/\textbf{92.10\%}  &  \textbf{100.00\%}/13.15\% &  \textbf{100.00\%}/23.68\%  \\
        CWE-416 &  \textbf{97.43\%}/\textbf{100.00\%} &  91.67\%/57.89\%  &    0.00\%/0.00\%   \\
        \bottomrule
    \end{tabular}
\end{table*}

For CWE-401, \toolname produces the 100\% recall on the ``maybe'' setting, outperforming all other tools.
While CodeQL has the highest precision (96.96\%), it misses almost half the defects.


For CWE-415, we are still doing well in terms of recall rate (92.10\%) while the highest percent of other tools is 23.68\%, but their accuracy reaches 100\% which is a little higher than ours.
At last for CWE-416, we perform well both in terms of accuracy(97.43\%) and recall rate (100\%), significantly higher than the others.

In summary, we believe \toolname achieves better results on precision and recall rate on Juliet test suites compared with the other tools.
The main reason that our tool outperforms is the precise control flow and the ability to analyze data flow especially across functions and files.

\subsection{RQ2: Is the declarative DSL easy to use?}

In \toolname, we implemented DSL by building the Fluent API using Java, as mentioned in section \ref{section:dsl}.
To highlight the project's ease of use benefits, we compare the DSL used for the CWEs described in section \ref{rq1}.

\begin{table}
\centering
\caption{DSL average line count on common CWE}
\label{tab:dsl-count}
\vspace{-6pt}
\begin{tabular}{c|ccc}\toprule
    \diagbox[width=10em]{CWE Type}{Tool Name} & \toolname & Joern & CodeQL \\
    \hline
    CWE-401 & 17   & 11    & 80     \\
    CWE-415 & 16   & 12    & 37     \\
    CWE-416 & 16   & 21    & 142    \\\hline
    AVG     & 16.3 & 14.7  & 86.3   \\\bottomrule
\end{tabular}
\vspace{-6pt}
\end{table}

We use existing queries of Joern \footnote{\url{https://queries.joern.io/}} and CodeQL \footnote{\url{https://github.com/github/codeql/blob/main/cpp/ql/src/Criticall}}, and write new queries for \toolname.
After removing blanks, comments, and imports, we compared the line counts for these DSL queries of each tool, and the result is shown in Table \ref{tab:dsl-count}.
On average, \toolname implements query rules using 16.3 lines of code, comparing 14.7 of Joern and 86.3 of CodeQL.
The statistics indicate that our DSL is much simpler and more concise compared to CodeQL and slightly more verbose than Joern.
However, compared to the high dependency on scripting language (Scala) of Joern, our declarative DSL is more user-friendly and much easier to write.

\subsection{RQ3: What is the efficiency of CPG extraction?}

As the scale of the project increases, it becomes more difficult to analyze the source code. 
Since queries can be continuously optimized through caching, we only focus on the performance of CPG extraction.
By converting C/C++ code to the granularity of statement level, QVoG has a clear advantage over Joern in memory consumption.
To adequately evaluate complex projects, we selected projects with line numbers from 10,000 to more than 1,500,000.
We run \toolname, Joern, and CodeQL on these projects respectively five times and measure the average time and memory cost of each one.
The result is shown in Table \ref{tab:TimeCost}.

\begin{table*}
    \caption{Time and Memory cost on projects of different scales}
    \label{tab:TimeCost}
    \begin{tabular}{c|cccc}\toprule
        \diagbox[width=10em]{Project/LOC}{Tool Name} & \toolname & Joern & CodeQL \\
        \hline
        zip\footnotemark{} 10,000+  &  \textbf{15s}/\textbf{160MB}  &  \textbf{15s}/300MB  &   6m30s/200MB  \\
        Django\footnotemark{} 100,000+ & \textbf{1m10s}/\textbf{200MB} & \textbf{1min10s}/2400MB & 1min30s/520MB &  \\
        Redis\footnotemark{} 150,000+ &  7min20s/2200MB  &  \textbf{2min}/2600MB &  3min40 (WSL)/\textbf{660MB}  \\
        SQLite\footnotemark{} 300,000+ &  12min20s/\textbf{780MB} &  \textbf{1min25s}/2300MB  &  7min/880MB   \\
    PostgreSQL\footnotemark{} 1,500,000+ &  15min/5200MB &  \textbf{7min}/6250MB &    19min (WSL)/\textbf{1200MB} \\ \bottomrule
    \end{tabular}
\end{table*}
\addtocounter{footnote}{-5}
\stepcounter{footnote}\footnotetext{\url{https://github.com/kuba--/zip}}
\stepcounter{footnote}\footnotetext{\url{https://github.com/django/django/releases/tag/3.0.2}}
\stepcounter{footnote}\footnotetext{\url{https://github.com/redis/redis/releases/tag/7.2.4}}
\stepcounter{footnote}\footnotetext{\url{https://www.sqlite.org/2024/sqlite-amalgamation-3450300.zip}}
\stepcounter{footnote}\footnotetext{\url{https://www.postgresql.org/ftp/source/v16.2/}}

From the statistics, we can see that \toolname approximately maintains a log increase in time complexity as the volume of code grows, which demonstrates good scalability.
Compared with other tools, \toolname has shown a commendable performance on both small-scale projects and extremely complex large projects. 
In terms of memory consumption, \toolname outperforms Joern, and in projects that are easier to build, i.e. require fewer cross-file analyses, we also outperform CodeQL.

For small-scale projects, our tool only takes 10 to 15 seconds as fast as Joern, outperforming CodeQL (6min30s).
Besides, we have the lowest memory cost (160MB).
By the way, in the first project with over 10,000 lines of code in Table \ref{tab:TimeCost}, the number of nodes in CPG of Joern reaches over 52,000 and the number of edges reaches over 472,000, both almost ten times that of our tool."

For extremely complex large projects, \toolname outperforms Joern in terms of memory consumption (almost 1G less) and outperforms CodeQL in terms of time consumption.

However, \toolname underperforms when the project contains extremely large source files.
For example, the source code for SQLite we used is the amalgamated version recommended by the official which consists of only four files.
In this case, the efficiency of \toolname is affected as we don't yet support intra-file paralleling.

\subsection{RQ4: Can machine learning alleviate the issues caused by over-specific rules?}

To address the issue of over-specific query rules, we trained two models to intelligently identify Sources and Sinks.
The training results are as follows.

\begin{table}[H]
\centering
\vspace{-1.0em}
\caption{$Model_{type}$ training result}
\vspace{-6pt}
\label{tab:type-model-train-result}
\begin{tabular}{cc} 
\hline
\textbf{Training Set Size}         & \textbf{Test Set Size}   \\ 
7328                      & 1832  \\ 
\hline
\textbf{Pre-Training Acc.}        & \textbf{Post-Training Acc.}  \\ 
26\%                    & 92\%  \\ 
\hline
\textbf{Pre Kappa Score} & \textbf{Post Kappa Score}  \\
0.004 & 0.89 \\\bottomrule
\end{tabular}
\end{table}

The results for $Model_{type}$ are shown in Table \ref{tab:type-model-train-result}.
From the result, we can see a significant increase in accuracy of the $Model_{type}$ from 26\% to 92\%, along with a Kappa coefficient of 0.89.
It demonstrates the excellence of the training results, and the high Kappa value indicates a very strong agreement between the predicted and actual classifications.

\begin{table}[H]
\centering
\vspace{-1.0em}
\caption{$Model_{pair}$ training result}
\vspace{-6pt}
\label{tab:pair-model-train-result}
\begin{tabular}{cc} 
\hline
\textbf{Training Set Size}         & \textbf{Test Set Size}   \\ 
3748                      & 937  \\ 
\hline
\textbf{Pre-Training Acc.}        & \textbf{Post-Training Acc.} \\ 
20\%                    & 92\%  \\ 
\hline
\textbf{Pre F1 Score.}        & \textbf{Post F1 Score.} \\
0.31 & 0.84 \\\bottomrule
\end{tabular}
\end{table}

The results for $Model_{pair}$ are shown in Table \ref{tab:pair-model-train-result}.
The positive samples for the $Model_{pair}$ are derived from existing pairs of Source and Sink nodes.
Negative samples are generated by randomly selecting two non-matching nodes for each Source/Sink.
We can find that the accuracy has improved dramatically from an untrained 20\% to 92\%, with an F1 score reaching 0.84.
This demonstrates the model's strong classification capability, effectively discerning whether a given Source and Sink are paired.

We further adopted a case study approach to test the model's capability to resolve existing issues.
We examined a vulnerability, CVE-2023-50447, which affects the Pillow-10.1.0 project, specifically in the \lstinline{src/PIL/ImageMath.py} file within the \lstinline{eval()} function. 
This vulnerability allows an attacker to pass specific environmental parameters to \lstinline{PIL.ImageMath.eval()}, enabling arbitrary code execution.

We applied machine learning module of \toolname to scan Pillow-10.1.0, which successfully identified CVE-2023-50447. 
To benchmark against other tools, we also employed Joern and CodeQL to detect the same issue. 
CodeQL, in this case, failed to identify the vulnerability, whereas Joern requires a tailored query to uncover it.
However, crafting the appropriate query before the vulnerability's disclosure is inherently challenging.
This indicates the machine learning component help generalize the detection rule and find the defect variants.
However, it also flagged some additional vulnerabilities that were of no vulnerability issue.
Addressing and reducing these false positives is a key priority in our upcoming work.


\section{Related Works}

In this section, we discuss the common static analysis approaches with a focus on graph-based analysis and machine learning.

\subsection{Static Analysis}

Traditional static code analysis techniques have evolved to include methods based on code similarity, symbolic execution, and rule-based detection.

Detection based on code similarity finds vulnerabilities in code by comparing it with known defective code.
Early research only includes lexical analysis of source code which lacks a deeper understanding of the syntax.
Scandariato et al. \cite{Scandariato2014} extracted terms and frequencies from source code to predict vulnerabilities in software components.
Yamaguchi et al. \cite{266592} use API sequence to represent code behavior to find possible defects.
To have a better understanding of the semantics of the code, researchers began to use advanced representations like trees and graphs.
For instance, SecureSync \cite{Pham2010} employs AST and GrouMiner \cite{Nguyen2009} uses CFG for vulnerability detection.

Symbolic execution finds potential security vulnerabilities by systematically traversing and analyzing the execution paths of programs. 
For example, Liang et al. \cite{Liang2016} analyzed the source code based on LLVM and used symbolic execution to detect program defects.
The method based on symbolic execution can determine the trigger of the vulnerability, but it also suffers from high overhead and may lead to path explosion.

Rule-based approaches use a set of pre-defined patterns to identify vulnerabilities, which are often extracted from known defects.
This approach is good at detecting well-defined vulnerabilities but may miss more general problems and produce false positives. \cite{Udrea2008}

\subsection{Graph-based Analysis}

With the continuous development of code defect detection, the graph has been regarded as a more expressive representation.
In this way, defect detection and analysis could be done by performing traversals on graphs.

Firstly, researchers proposed a vulnerability detection method based on the AST \cite{Yamaguchi2012}.
By analyzing the AST, this method enables known vulnerabilities to be broken down into code with similar structure characteristics.
This approach was evaluated in four popular open-source projects and successfully found zero-day vulnerabilities.
Later on, analysis based on Control Flow Graph \cite{Halim2019} and Program Dependency Graph \cite{Ferrante1987} emerged to better utilize flow information within code.
By combining graphs, Yamaguchi et al. \cite{Yamaguchi2014} proposed a novel representation of Code Property Graph.
It consists of AST, CPG, and PDG thus providing more detailed information of the program.
For the industry scenario, Joern and CodeQL are two renowned examples.
Joern uses an in-memory graph database to store CPG information while CodeQL uses a relational database.
Users can use the DSL provided by each tool to scan the codebase for vulnerabilities.
Although analysis based on graph analysis has made great progress in recent years, it still struggles with scalability and can produce many false positives and negatives.

\subsection{Deep Learning-based Analysis}

Deep learning has seen rapid development in code detection, with models like CNN, RNN, LSTM, Transformer, and GNN being applied. 
Using deep learning, Fu et al. \cite{Fu2022} designed a binary classification task based on Transformer to identify the lines of vulnerabilities through attention score.
Later, Fu et al. \cite{Fu2023} also proposed a hierarchical neural Transformer network to distill vulnerability knowledge.
In addition, Wang et al. \cite{Xiaomeng2018} proposed an end-to-end code inspection framework combining NLP and CPG.
It made significant improvements in precision, and recall rate compared with traditional approaches.

\section{Conclusion}

Static analysis can help detect defects and vulnerabilities in the early stages of software development.
Code Property Graph, as a novel representation of source code, enables vulnerability detection via graph queries.
In this paper, we introduced \toolname as a general approach for defect and vulnerability detection.
It performs analysis based on a compressed CPG to eliminate redundancy and improve query efficiency.
To fully support vulnerability queries, we also designed a declarative DSL along with a set of query libraries.
Furthermore, we integrated machine learning to tackle the overfitting problem of specific vulnerability patterns.
Compared with existing renowned tools Joern and CodeQL, \toolname shows satisfying performance on CPG extraction.
For common CWE vulnerabilities, our tool has a higher precision and recall rate on average and demonstrates a better generalization ability with machine learning.

\bibliographystyle{unsrt}
\bibliography{references}

\end{document}